\pdfoutput=1
\documentclass[a4wide,11pt,bezier]{article}
\usepackage[margin=3cm]{geometry}
\usepackage{amsmath}
\usepackage[noblocks,affil-it]{authblk}
\usepackage[pdftex]{graphicx,color}
\usepackage[small]{caption}
\usepackage[htt]{hyphenat}
\usepackage{slashed}
\usepackage{hyperref}
\usepackage{booktabs}
\title{\texttt{CAMORRA}: a C++ library for recursive computation of particle scattering amplitudes}
\author{R. Kleiss}
\affil{IMAPP, Radboud Universiteit Nijmegen, Huygens Building, Heyendaalsweg 135, 6525AJ Nijmegen, Netherlands}
\author{G. van den Oord}
\affil{NIKHEF, Kruislaan 409, 1098SJ Amsterdam, Netherlands}
\begin{document}

\maketitle

\begin{abstract}
We present a new Monte Carlo tool that computes full tree-level matrix elements in high-energy physics. The program accepts user-defined models and has no restrictions on the process multiplicity. To achieve acceptable performance, \texttt{CAMORRA} evaluates the matrix elements in a recursive way by combining off-shell currents. Furthermore, \texttt{CAMORRA} can be used to compute amplitudes involving continuous color and helicity final states.
\end{abstract}

\section{Introduction}
In the present LHC- and Tevatron-physics dominated era, the need for understanding of processes with several hard final-state partons has emerged. As the number of Feynman diagrams grows factorially with the final state multiplicity, new recursive methods have been developed \cite{BerendsNPB306,CaravagliosPLB358} and implemented \cite{ALPGEN,HELAC,COMIX}. The recursive method essentially factorizes the amplitude into contributing off-shell currents, which are stored in the computer's memory, so that the number of operations grows only exponentially. Moreover the method allows a fully numerical approach, where an initialization run determines which currents can be combined to a higher-level one, and the matrix element is evaluated by applying the recursive relations in the vertex tree. The input from the physics model is coded in a set of recursive relations, which are trivially constructed from the Feynman rules. Hence there is no reason why the recursive method should be limited to standard-model amplitudes, especially since many new physics signatures are associated with multi-jet events or long decay chains at the LHC. Full new-physics matrix elements, treating signal and background on equal footing, seem indispensable to the BSM phenomenologist's toolbox.\\
\\
The purpose of \texttt{CAMORRA}\footnote{Abbreviation of CAravaglios-MORetti Recursive Algorithm, freely available at http://www.nikhef.nl/$\sim$vdoord} is to provide a stand-alone, user-friendly, modular and model-independent implementation of the recursive algorithm. As such, the library can serve as a matrix-element computing engine for a Monte-Carlo generator, or be used to reweigh an existing sample of events. For the latter purpose a Les-Houches event \cite{LHEF} interface is provided, which reads events sequentially and automatically builds the subprocess vertex trees. To achieve a performance competitive with similar Fortran codes, the library relies heavily on \emph{template} techniques, allowing the compiler to inline the recursive relations and optimize the resulting machine code. Furthermore, multiplications by constant, sparse matrices such as the Dirac matrices are coded by hand and dynamic binding is avoided wherever possible.\\
\\
Together with the core routines, various ready-to-use models are included in the package, as well as a flat phase space generator, and generators for color and helicity configurations.
\begin{figure}
\begin{center}
\resizebox{12cm}{!}{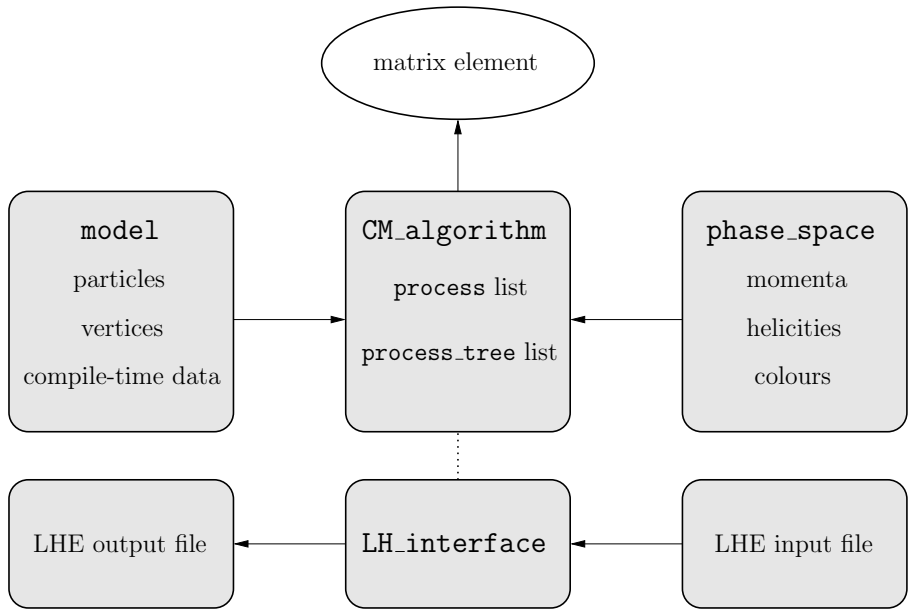}
\end{center}
\caption{Outline of Camorra's basic flow structure.}
\label{flowchart}
\end{figure}
\section{Program Outline}
A rough sketch of the library is given by the flow chart in Fig. \ref{flowchart}. All classes and functions of the library reside in the namespace \texttt{Camorra}, and all header files will be installed in a subdirectory named Camorra, so a program using the library will typically begin as follows:\\
\\
\indent\texttt{\#include<Camorra/CM\_algorithm.h>}\\
\indent\texttt{using namespace Camorra;}\\
\subsection{Algorithm class}
The main class which distributes a computation is called \texttt{CM\_algorithm}. It takes the model type and incoming and outgoing particles as template parameters, which are therefore compile-time constants and common to all subprocesses. Its constructor and process-addition method take character string arguments, e.g.\\
\\
\indent\texttt{CM\_algorithm<QCD,2,3>myalgo("g,g > t,tbar,g");}\\
\indent\texttt{myalgo.add\_process("u,ubar > t,tbar,g");}\\
\\
At construction, the algorithm class will first create a (unique) instance of the model, and a list of all possible currents for the process multiplicity. This list is static, which means that all subprocesses and algorithm instances of the same multiplicity in the program will make use of it. Furthermore it decomposes the subprocess character strings and matches these with particle names, creating a list of \texttt{process} objects. This list is ordered for efficient lookup and double-counted processes are erased using the algorithm's \texttt{load()} method. Calling\\
\\
\indent\texttt{myalgo.construct\_trees();}\\
\\
sets up vertex trees for the subprocesses. This routine consecutively performs the following actions:
\begin{itemize}
\item Choose a final-current particle (by default the first one, but this can be manually controlled) and set up the zero-level currents.
\item Recursively recombine currents into higher-level ones according the the model's vertex content.
\item When the highest-level currents are constructed, attempt to match with the final current. Recursively run down the tree, deleting all vertices combining to non-matching highest-level currents.
\item Determine Fermi signs and fermion flow reversal flags. 
\end{itemize}
The algorithm also contains an iterator running over the process list. It is controlled by the algorithm's \texttt{set\_process} member function, taking a character string argument of the same form as before. For faster lookup one may use \texttt{set\_process\_flavours} or \texttt{set\_process\_pdg\_ids}, which take integer vectors as arguments. Note that if the process is not found, \texttt{CAMORRA} will attempt to insert it into the list and construct a vertex tree on the fly. Once a subprocess is selected, the user attains access to the phase space instances of the external particles by calling \texttt{myalgo.get\_phase\_space(i)}. This interface (see appendix \ref{PSinterface}) depends on some compile-time data stored in the model class, which will be explained in greater detail in the upcoming section. The evaluation of the amplitude proceeds by calling\\
\\
\indent\texttt{myalgo.evaluate();}\\
\\
The evaluation first cleans the list of currents, setting all participating sub-amplitudes to zero. Then it runs over its list of vertices, applying the series of recursive relations to the off-shell currents. Finally it returns the contraction between the highest-level current and final one. The algorithm class also provides a spin-summation version of the evaluation function called \texttt{evaluate\_spin\_sum()}. A helicity summation flag for external particle $i$ is set by \texttt{myalgo.sum\_spin(i)}, the collective flag is set by calling the \texttt{sum\_spins()} member.
\subsection{LH interface class}
Interfacing with other Monte Carlo tools via the Les Houches standard format is achieved by constructing a object of the \texttt{LH\_interface} type. This is a daughter class of \texttt{CM\_algorithm}, with extra methods for reading and writing Les Houches Accord event files. Les Houches input starts with the initialization block readout via the member function \texttt{read(}\emph{filename.lhe}\texttt{)}, followed by the sequential event block readout functions \texttt{read\_event()}, \texttt{find\_process()} and \texttt{input\_event()}. The first copies the data into a memory buffer and returns a positive boolean if the event is compatible with the algorithm, the second searches the process list and occasionally creates a new tree and the third copies the momenta and colors to the tree. If some masses in the event file do not match those in the model file (effective quark masses, for example), the member function \texttt{rescale\_phase\_space()} rescales the rest-frame 3-momenta such that the invariant masses match the ones specified in the model. Output to LHE files is trivially obtained by calling the \texttt{write(}\emph{filename.lhe}\texttt{)} method at initialization and the \texttt{write\_event()} function subsequently.

\section{Model implementation}
The specialty of the \texttt{CAMORRA} library is the ability to deal with a broad range of user-defined models. The current version supports theories containing scalar and vector particles and fermions. For the latter both Dirac and Majorana representations are allowed. On the vertex side, \texttt{CAMORRA} supports a range of built-in Lorentz and color structures which can be arbitrarily combined. The modular structure of the package makes it possible to construct new structures without modifying any of the existing code. The vertex multiplicity is however limited to four. Effective theories containing higher rank vertices can be included by introducing auxiliary fields, which are also supported by the library. Furthermore it should be noted that all parameters in a model are passed by reference and hence can be modified at runtime without having to invoke any function updating the recursive relations.
\subsection{Compile-time data}\label{modeldef}
To implement a new physics model, one derives a class from the \texttt{model<}\emph{derived model}\texttt{>} class template. The relevant data to \texttt{CAMORRA} contained in this class is separated into two groups: global, compile-time statements and a constructor consisting of a series of particle and vertex additions. The model's header file would for example look as follows,\\
\\
\texttt{class myBSM: public model<myBSM>}\\
\texttt{\{}\\
\indent\texttt{typedef double value\_type;}\\
\indent\texttt{typedef Minkowski\_type spacetime\_type;}\\
\indent\texttt{typedef Pauli\_basis Dirac\_algebra\_type;}\\
\indent\texttt{typedef helicity\_type spin\_vector\_type;}\\
\indent\texttt{typedef colour\_flow colour\_treatment;}\\
\indent\texttt{static const unsigned dimension=4;}\\
\indent\texttt{static const int beam\_direction=3;}\\
\indent\texttt{static const bool coloured=1;}\\
\indent\texttt{static const bool continuous\_helicities=1;}\\
\indent\texttt{static const bool continuous\_colours=0;}\\
\indent\texttt{myBSM();}\\
\texttt{\};}\\
\\
The first line represents the numerical type used throughout the computation of the matrix elements. One can choose any floating-point type for which the standard library's math functions are supported. The second and fifth line determine the space-time metric used throughout the program. Choosing \texttt{Euclidean\_type} compiles all vector contractions to Euclidean inner products. However, in the current version the spinor and vector wave functions are only defined in four-dimensional Minkowski space. The second line chooses the basis of gamma matrices used in \texttt{CAMORRA}. This option is mainly implemented for checking purposes, although changing it to \texttt{Weyl\_basis} may yield a slight speedup. Together with the beam direction definition it gives the algorithm enough information to construct polarization vectors and massless spinors. Massive spinor wave functions however require a definition of the spin vector type--given here on line three--which represents the choice for the construction of a spin vector appearing in the massive spinor formulas (see section \ref{pols}). The boolean \texttt{coloured} should be defined and  set positive whenever the model contains multiplets of (unbroken) symmetries. If it is set false, \texttt{CAMORRA} runs in colorless mode, and no definition of \texttt{colour\_treatment} is necessary. If true, the program needs this information for the treatment of objects transforming under the $SU(N)$ adjoint representation. Choosing \texttt{adjoint} treats gluons as octets and all color structures accordingly, while for the choice above the gluons are represented as $q\bar{q}$ pairs in color space and all color Feynman rules are expanded in this basis \cite{MaltoniPRD67}. Finally there are the continuous color and helicity flags, controlling the phase space integration method. These values determine the interface to the phase space instances of external particles in a process (for more details, see appendix \ref{PSinterface}).
\subsection{Particle insertion}
The final line in the code snippet above declares the constructor of the model. Its implementation should consist of a series of statements that define the particle and vertex content of the theory. Particle insertion methods are\\
\\
\indent\texttt{add\_scalar<}\emph{representation}\texttt{>(}\emph{argument list}\texttt{);}\\
\indent\texttt{add\_fermion<}\emph{representation}\texttt{>(}\emph{argument list}\texttt{);}\\
\indent\texttt{add\_vector<}\emph{representation,gauge}\texttt{>(}\emph{argument list}\texttt{);}\\
\\
Here \emph{representation} is an optional template parameter type denoting the color degrees of freedom of the field. Currently \texttt{fundamental\_rep< SU<N> >} or \texttt{adjoint\_rep< SU<N> >} are supported. Representation types may be chained using the \texttt{compose<>} class template to define particles with multiple colors. The \emph{gauge} parameter in the vector inserter function can be either \texttt{Feynman\_gauge}, \texttt{unitary\_gauge} or \texttt{R\_vector\_gauge}. The argument list consist of a character string denoting the name of the particle, pointers to the mass and width and an integer denoting the particle data group code. Except for the name, any of these fields may be omitted to yield zero, e.g.:
\begin{itemize}
\item\texttt{add\_vector<Feynman\_gauge>("gamma",22)}: insert a massless vector in the Feynman gauge with pdg id 22,
\item\texttt{add\_fermion< adjoint\_rep< SU<3> > >("}$\sim$\texttt{g",\&m,\&w)}: insert a (Majorana) fermion octet with mass \texttt{m} and width \texttt{w}.
\end{itemize}
Particle-anti-particle pairs are inserted with the corresponding functions \texttt{add\_scalars}, \texttt{add\_fermions}, \texttt{add\_vectors}, which take two names in their argument list, e.g.\\
\\
\texttt{add\_fermions< fundamental\_rep< SU<3> > >("b","bbar",\&mb,5);}
\subsection{Vertex insertion}
After the series of particle definitions, the vertex insertion functions can be used to define interactions. These are of the form\\
\\
\texttt{add\_vertex<}\emph{color structure,spacetime structure}\texttt{>(}\emph{particles,couplings}\texttt{)};\\
\\
where the argument list contains three or four particle names and the couplings are pointers to complex floating-point numbers. Usually only one coupling is allowed, except for the V-A type Lorentz structures which require two (see table \ref{Feynruletable}). Some enlightening examples are
\begin{itemize}
\item \texttt{add\_vertex<vff>("gamma","e+","e-",\&e)}: inserts a $\gamma^{\mu}$-type coupling in QED with coupling constant stored in the variable \texttt{e}.
\item \texttt{add\_vertex<colour\_tensor::d<}\emph{triplet}\texttt{,1,2>,vffVA>("Z",ubar","u",\&cV, \&cA)}: inserts a coupling of the form
\begin{equation*}
\delta_{\alpha\beta}\gamma^{\mu}(c_V+c_A\gamma^5)\,,
\end{equation*}
where the values $c_V$ and $c_A$ have been assigned to the variables \texttt{cV} and \texttt{cA}. In the color structure we have substituted \texttt{fundamental\_rep< SU<3> >} by \emph{triplet}. Note that all color structures in table \ref{coltable} reside in the namespace \texttt{colour\_tensor} and depend on template parameters themselves. For the Kronecker delta \texttt{d}, the first one is the representation type of the contracted indices, the second and third denote the fields carrying these color indices.
\item The squark-quark-gluino vertex in supersymmetric QCD,\\
\begin{tabular}{cc}
\includegraphics[width=5.5 cm]{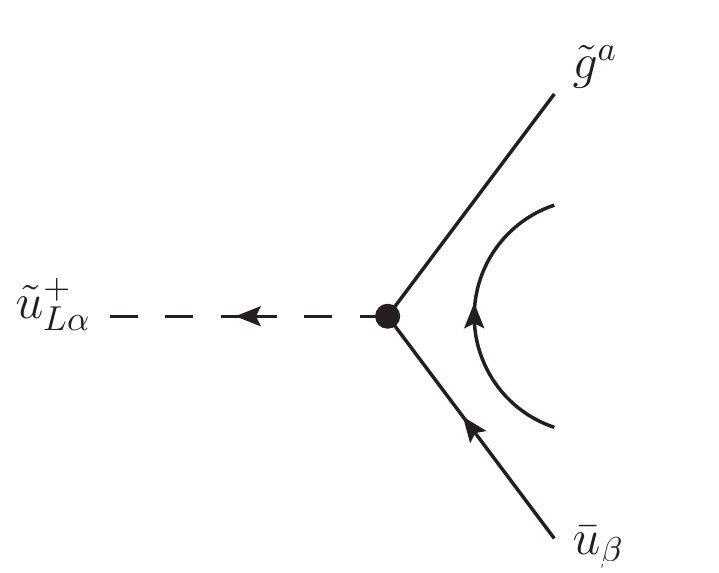}&\raisebox{12ex}{$\qquad gT^a_{\alpha\beta}(1-\gamma^5)$}\\
\end{tabular}\\
has a nontrivial ordering of the color indices, as the default ordering of the $SU(N)$ generator \texttt{colour\_tensor::T<}\emph{triplet}\texttt{,0,1,2>} assumes the first particle to be in the adjoint representation and the second and third in the (anti)-fundamental representations (see table \ref{coltable}). The insertion code for the above vertex looks like (omitting the color tensor namespace for brevity)\\
\\
\texttt{add\_vertex<T<}\emph{triplet}\texttt{,2,0,1>,sffR>("}$\sim$\texttt{u\_L+","ubar","}$\sim$\texttt{g");}\\
\end{itemize}
During construction of the vertices, \texttt{CAMORRA} checks if the indices of the particles in the argument list match those of the Feynman rule parameter. If this fails, the vertex is omitted from the internal list and an error message is issued in the log file. Note however that the program checks only if index \emph{ranges} are in correspondence with the recursive equations, so that for example a vector particle appearing in a spinor entry of a vertex inserter will not be noticed (at least, in four dimensions). Again there is the option to compose color structures for vertices with multiply-colored legs with the template type \texttt{compose<...>}.
\begin{table}
\begin{tabular*}{\textwidth}{@{\extracolsep{\fill}}ccccc}
\toprule
header file & vertex class & arguments & couplings & Feynman rule\\
\midrule
\texttt{sss.h} & \texttt{sss} & $S_1,S_2,S_3$ & 1 & $c_1$\\
\texttt{ssss.h} & \texttt{ssss} & $S_1,S_2,S_3,S_4$ & 1 & $c_1$\\
\texttt{svv.h} & \texttt{svv} & $S_1,V_2^{\mu},V_3^{\nu}$ & 1 & $c_1g^{\mu\nu}$\\
\texttt{ssvv.h} & \texttt{ssvv} & $S_1,S_2,V_3^{\mu},V_4^{\nu}$ & 1 & $c_1g^{\mu\nu}$\\
\texttt{vss.h} & \texttt{vss} & $V_1^{\mu},S_2,S_3$ & 1 & $c_1(p_2-p_3)^{\mu}$\\
\texttt{vvv.h} & \texttt{vvv} & $V_1^{\mu},V_2^{\nu},V_3^{\rho}$ & 1 & $c_1\Big(g^{\mu\nu}(p_1-p_2)^{\rho}+g^{\mu\rho}(p_3-p_1)^{\nu}$\\
&&&&$+g^{\nu\rho}(p_2-p_3)^{\mu}\Big)$\\
\texttt{vvvv.h} & \texttt{vvvv} & $V_1^{\mu},V_2^{\nu},V_3^{\rho},V_4^{\sigma}$ & 1 & $c_1(2g^{\mu\rho}g^{\nu\sigma}-g^{\mu\nu}g^{\rho\sigma}-g^{\mu\sigma}g^{\nu\rho})$\\
\texttt{sff.h} & \texttt{sff} & $S_1,F_2^{\alpha},F_3^{\beta}$ & 1 & $c_1\delta^{\alpha\beta}$\\
\texttt{sff5.h} & \texttt{sff5} & $S_1,F_2^{\alpha},F_3^{\beta}$ & 1 & $c_1\gamma_5^{\alpha\beta}$\\
\texttt{sffL.h} & \texttt{sffL} & $S_1,F_2^{\alpha},F_3^{\beta}$ & 1 & $\tfrac{1}{2}c_1(1+\gamma_5)^{\alpha\beta}$\\
\texttt{sffR.h} & \texttt{sffR} & $S_1,F_2^{\alpha},F_3^{\beta}$ & 1 & $\tfrac{1}{2}c_1(1-\gamma_5)^{\alpha\beta}$\\
\texttt{sffVA.h} & \texttt{sffVA} & $S_1,F_2^{\alpha},F_3^{\beta}$ & 2 & $(c_1+c_2\gamma_5)^{\alpha\beta}$\\
\texttt{sffLR.h} & \texttt{sffLR} & $S_1,F_2^{\alpha},F_3^{\beta}$ & 2 & $\tfrac{1}{2}(c_1+c_2+(c_1-c_2)\gamma_5)^{\alpha\beta}$\\
\texttt{vff.h} & \texttt{vff} & $V_1^{\mu},F_2^{\alpha},F_3^{\beta}$ & 1 & $c_1(\gamma^{\mu})^{\alpha\beta}$\\
\texttt{vff5.h} & \texttt{vff5} & $V_1^{\mu},F_2^{\alpha},F_3^{\beta}$ & 1 & $c_1(\gamma_5\gamma^{\mu})^{\alpha\beta}$\\
\texttt{vffL.h} & \texttt{vffL} & $V_1^{\mu},F_2^{\alpha},F_3^{\beta}$ & 1 & $\tfrac{1}{2}c_1((1+\gamma_5)\gamma^{\mu})^{\alpha\beta}$\\
\texttt{vffR.h} & \texttt{vffR} & $V_1^{\mu},F_2^{\alpha},F_3^{\beta}$ & 1 & $\tfrac{1}{2}c_1((1-\gamma_5)\gamma^{\mu})^{\alpha\beta}$\\
\texttt{vffVA.h} & \texttt{vffVA} & $V_1^{\mu},F_2^{\alpha},F_3^{\beta}$ & 2 & $((c_1+c_2\gamma_5)\gamma^{\mu})^{\alpha\beta}$\\
\texttt{vffLR.h} & \texttt{vffLR} & $V_1^{\mu},F_2^{\alpha},F_3^{\beta}$ & 2 & $\tfrac{1}{2}((c_1+c_2+(c_1-c_2)\gamma_5)\gamma^{\mu})^{\alpha\beta}$\\
\texttt{symtvv.h} & \texttt{symtvv} & $T_1^{\mu\nu},V_2^{\rho},V_3^{\sigma}$ & 1 & $c_1(g^{\mu\rho}g^{\nu\sigma}+g^{\mu\sigma}g^{\nu\rho})$\\
\texttt{asymtvv.h} & \texttt{asymtvv} & $T_1^{\mu\nu},V_2^{\rho},V_3^{\sigma}$ & 1 & $c_1(g^{\mu\rho}g^{\nu\sigma}-g^{\mu\sigma}g^{\nu\rho})$\\
\bottomrule
\end{tabular*}
\caption{Available Feynman rules in the package. The third column denotes the types of the particle arguments in the \texttt{add\_vertex<}\emph{vertex class}\texttt{>(...)} function, where $S$ denotes a scalar, $V$ a vector and $F$ a fermion type. The number of coupling constants is listed in the fourth column, represented by $c_1,c_2,\ldots$ in the last column.}
\label{Feynruletable}
\end{table}
\begin{table}[h]
\begin{tabular*}{\textwidth}{@{\extracolsep{\fill}}cccc}
\toprule
header file & tensor class & arguments & Color factor\\
\midrule
\texttt{d.h} & \texttt{d<R}$,i,j$\texttt{>} & $\{\phi_{i\alpha},\phi_{j\beta}\}$ & $\delta_{\alpha\beta}$\\
\texttt{dd.h} & \texttt{dd<R}$,i,j,k,l$\texttt{>} & $\{\phi_{i\alpha},\phi_{j\beta},\phi_{k\gamma}\phi_{l\delta}\}$ & $\delta_{\alpha\gamma}\delta_{\beta\delta}$\\
\texttt{dd\_plus.h} & \texttt{dd\_plus<R}$,i,j,k,l$\texttt{>} & $\{\phi_{i\alpha},\phi_{j\beta},\phi_{k\gamma}\phi_{l\delta}\}$ & $\delta_{\alpha\gamma}\delta_{\beta\delta}+\delta_{\alpha\delta}\delta_{\beta\gamma}$\\
\texttt{dd\_min.h} & \texttt{dd\_min<R}$,i,j,k,l$\texttt{>} & $\{\phi_{i\alpha},\phi_{j\beta},\phi_{k\gamma}\phi_{l\delta}\}$ & $\delta_{\alpha\gamma}\delta_{\beta\delta}-\delta_{\alpha\delta}\delta_{\beta\gamma}$\\
\texttt{f.h} & \texttt{f<G}$,i,j,k$\texttt{>} & $\{\phi_i^a,\phi_j^b,\phi_k^c\}$ & $f^{abc}$\\
\texttt{ff\_contr.h} & \texttt{ff\_contr<G}$,i,j,k,l$\texttt{>} & $\{\phi_i^a,\phi_j^b,\phi_k^c,\phi_l^d\}$ & $f^{abe}f_e^{\phantom{e}cd}$\\
\texttt{T.h} & \texttt{T<R}$,i,j,k$\texttt{>} & $\{\phi_i^a,\phi_{j\alpha},\phi_{k\beta}\}$ & $T^a_{\phantom{a}\alpha\beta}$\\
\texttt{TT.h} & \texttt{TT<R}$,i,j,k,l$\texttt{>} & $\{\phi_i^a,\phi_j^b,\phi_{k\alpha},\phi_{l\beta}\}$ & $T^a_{\phantom{a}\alpha\gamma}T^{b\gamma}_{\phantom{b\gamma}\beta}$\\
\texttt{TT\_plus.h} & \texttt{TT\_plus<R}$,i,j,k,l$\texttt{>} & $\{\phi_i^a,\phi_j^b,\phi_{k\alpha},\phi_{l\beta}\}$ & $T^a_{\phantom{a}\alpha\gamma}T^{b\gamma}_{\phantom{b\gamma}\beta} +T^b_{\phantom{a}\alpha\gamma}T^{a\gamma}_{\phantom{b\gamma}\beta}$\\
\texttt{TT\_min.h} & \texttt{TT\_min<R}$,i,j,k,l$\texttt{>} & $\{\phi_i^a,\phi_j^b,\phi_{k\alpha},\phi_{l\beta}\}$ & $T^a_{\phantom{a}\alpha\gamma}T^{b\gamma}_{\phantom{b\gamma}\beta} -T^b_{\phantom{a}\alpha\gamma}T^{a\gamma}_{\phantom{b\gamma}\beta}$\\
\texttt{TT\_contr.h} & \texttt{TT\_contr<R}$,i,j,k,l$\texttt{>} & $\{\phi_{i\alpha},\phi_{j\beta},\phi_{k\gamma},\phi_{l\delta}\}$ & $T^a_{\phantom{a}\alpha\beta}T^{}_{a\gamma\delta} $\\
\bottomrule
\end{tabular*}
\caption{Available color structures in the package. The first template argument of the color tensor class is either a group type (\texttt{G}) or a representation type (\texttt{R}). The other arguments are distinct integers $i,j,\ldots$ which should be smaller than the rank of the vertex, and which determine the ordering of the multiplet argument list in the third column. In the most right column the adjoint-representation indices are labeled $a,b,\ldots$ and other representation indices by $\alpha,\beta,\ldots$.}
\label{coltable}
\end{table}
\subsection{Additional features}
Finally there are some extra features in the model base class. First, there is a function that defines a particle family, which can serve as an argument for the algorithm class to construct a list of subprocesses in one go. For example, the particle content of a jet in the standard model is reflected by a statement in the \texttt{SM} constructor:\\
\\
\indent\texttt{construct\_family("j","u,ubar,d,dbar,c,cbar,s,sbar,b,bbar,g");}\\
\\
making the $j$ name available in multi-process algorithm instances like \texttt{CM\_algorithm<SM,2,2> ("e+,e- > j,j")}. Again, the \texttt{load()} method will ensure only unique processes are inserted in the list. Secondly, a model inherits the \texttt{decouple\_particle()} and \texttt{decouple\_vertex()} functions. Because \emph{erasing} a particle or vertex from the global instance's lists is rather unsafe (previously built algorithm instances become invalid), the decoupling functions only set a flag to the particle or vertex, omitting their contributions from future amplitude evaluations.
\subsection{Implemented models}
There are several predefined models in the package. Some toy models for checking purposes are \texttt{phi3}, \texttt{phi34} and \texttt{WZM}, constituting respectively a $\phi^3$- and $(\phi^3+\phi^4)$-theory and the Wess-Zumino model \cite{WessPLB49}. More physically relevant models are
\begin{itemize}
\item\texttt{scalar\_QED}: scalar quantum electrodynamics with a single flavor of scalar electrons.
\item\texttt{QED}: contains three flavors of leptons, of which only the $\tau$ is massive by default.
\item\texttt{QCD}: contains six quark flavors, of which the $c,b$ and $t$ are massive by default. The static member function \texttt{set\_four\_gluon\_vertex()} allows to switch between an four-gluon vertex description of the Lagrangian and an auxiliary tensor field treatment.
\item\texttt{EWSM}: the colorless electroweak standard model implementation, with massive $c,b,t$ and $\tau$ and both $ud$- and $bc$-mixing.
\item\texttt{SM}: the full standard model, combining all features of QCD and the EWSM classes.
\item\texttt{susy\_QED,susy\_QCD}: supersymmetric quantum electro- and chromodynamics. The sparticles have by default the same mass as their standard model partners and zero decay width.
\end{itemize}
All of the above model classes contain the following public static members:
\begin{itemize}
\item\texttt{alpha}, \texttt{alpha\_s} and \texttt{QCD\_scale}: electromagnetic and strong couplings, and the scale at which $\alpha_s$ is evaluated. The former is constant and equals -1 in \texttt{(susy\_)QCD}, the latter two in the colorless models. The functions \texttt{set\_alpha(}$\ldots$\texttt{)}, \texttt{set\_alpha\_s(}$\ldots$\texttt{)}, and \texttt{set\_QCD\_scale(}$\ldots$\texttt{)} insert new values for the respective parameters and update vertex couplings depending on them.
\item\texttt{refresh\_couplings()}: calculates all couplings from the current values of the basic parameters. The basic parameters are the strong and electromagnetic couplings, and for the \texttt{EWSM} and \texttt{SM} classes the additional input values \texttt{G\_F} (Fermi's constant), \texttt{M\_Z} ($Z$ pole mass) and \texttt{M\_h0} (Higgs pole mass) using the conventions of \cite{DennerFP41}.
\item\texttt{refresh\_masses()}: although the values of the masses are passed by reference and recursive relations therefore always use the current value of a parameter, switching from a zero mass to a nonzero value is correctly updated by the algorithm once this function has been called. This is because the choice between the massless or massive polarization routines is not checked at each amplitude evaluation.
\item\texttt{set\_unitary\_gauge()}: replaces all massive and massless vector propagators respectively by
\begin{equation*}
\frac{-ig^{\mu\nu}+ip^{\mu}p^{\nu}/M^2}{p^2-M^2}\,,\qquad \frac{-ig^{\mu\nu}+ip^{\mu}p^{\nu}/p^2}{p^2}\,.
\end{equation*}
In the electroweak sector, the would-be Goldtone bosons are decoupled.
\item\texttt{set\_R\_xi\_gauge()}: replaces all vector and would-be Goldstone propagators respectively by
\begin{equation*}
\frac{-ig^{\mu\nu}+i(1-\xi)p^{\mu}p^{\nu}/(p^2-\xi M^2)}{p^2-M^2}\,,\qquad\frac{i}{p^2-\xi M^2}\,.
\end{equation*}
\end{itemize}
The (electroweak) standard model classes also provide built-in functions for computing (leading order) decay widths of the Higgs, top and $W$ and $Z$ bosons. These formulas assume all fermions massless except for the top and bottom quarks. Inclusion of finite widths is implemented by the complex mass scheme \cite{DennerNPB560}, absorbing the width in a complex mass through
\begin{equation*}
\tilde{M}=\sqrt{M^2-iM\Gamma}\,,
\end{equation*}
and performing the computation of the electroweak parameters with $\tilde{M}_Z,\tilde{M}_W$ and $\tilde{M}_h$ instead of the real pole masses. Because such a substitution preserves all Ward identities, resonant diagrams will still constitute gauge-invariant contributions to the amplitude.\\
\\
To simplify extensions all of the above models have templated counterparts called \texttt{QED\_base<}\emph{derived model,value type}\texttt{>}, \texttt{SM\_base<}\emph{derived model,value type}\texttt{>}, etc. Deriving a new model class from one of these automatically inserts their particle and vertex content in the user-defined class. Contrary to the \texttt{model} base class, these carry 2 template arguments. The second is the numerical type of the couplings and masses, and should be equal to the \texttt{value\_type} defined in the derived model.

\section{Library details}

\subsection{Polarizations}\label{pols}
For the construction of massless spinor wave functions, \texttt{CAMORRA} adopts the spinor techniques developed in \cite{KleissNPB241,KleissNPB262}. From  the \texttt{beam\_direction} value in the model class, the algorithm constructs two momenta $k_0$ and $k_1$ that fulfill 
\begin{equation*}
k_0^2=0\,,\qquad k_1^2=-1\,,\qquad k_0\cdot k_1=0
\end{equation*}
and for which it assumes no momenta will be parallel to. For the common choice 3 (beam direction along the $z$-axis), these will be chosen to be $k_0=(1,1,0,0)$ and $k_1=(0,0,1,0)$.
Then, the \texttt{Dirac\_algebra\_type} is used to construct two statically defined \emph{basic spinors} $u_{\pm}(k_0)$, fulfilling
\begin{equation*}
u_{\pm}(k_0)\bar{u}_{\pm}(k_0)=\tfrac{1}{2}(1\mp\gamma^5)\slashed{k}_0\,,\quad u_+(k_0)=\slashed{k}_1u_-(k_0)\,,
\end{equation*}
The massless fermion wave functions can be directly constructed from the constant spinors above. To cope with Majorana particles however, the relative phase between spinor and antispinors has to be fixed by the charge conjugation matrix, $u_{\lambda}(p)=C\bar{v}_{\lambda}^T(p)$ \cite{KleissEPJC64}.
The reason for this feature of the spinor phase choice will be explained in the next section. Hence, the algorithm will construct the positive-helicity spinors with the basic spinors, and the negative-helicity ones with the explicit charge conjugation:
\begin{equation*}
u_{+}(p)=\frac{1}{\sqrt{2p\cdot k_0}}\slashed{p}u_{-}(k_0)\,,\qquad u_-(p)=C\bar{u}^T_+(p)\,.
\end{equation*}
For the construction of massless polarization vectors, the beam direction constant is used once more to construct a third massless momentum $k_2$ not parallel to any momenta. For the standard choice \texttt{beam\_direction=3}, the choice $k_2=k_1+(1,0,0,0)$ is made and massless helicity vectors are constructed from
\begin{equation*}
\varepsilon_{\lambda}^{\phantom{\lambda}\mu}(p)=\frac{1}{\sqrt{4p\cdot k_2}}\bar{u}_{\lambda}(p)\gamma^{\mu}u_{\lambda}(k_2)
\end{equation*} 
The spacetime class also defines a function splitting a massive momentum $p$ into two massless vectors $p_1$ and $p_2$ whose spatial parts are (anti-)parallel to $\vec{p}$. The massive vector helicity states are then evaluated from
\begin{equation*}
\varepsilon_{\lambda}^{\phantom{\lambda}\mu}(p)=\frac{1}{\sqrt{2}m}\bar{u}_{\lambda}(p_1)\gamma^{\mu}u_{\lambda}(p_2)\,,\qquad 
\varepsilon_0^{\phantom{0}\mu}=\frac{p_1^{\mu}-p_2^{\mu}}{m}\,.
\end{equation*}
For the standard choices of Dirac algebra basis and beam direction the formulas above were evaluated symbolically using MAPLE and the resulting expressions coded by hand. For nonstandard bases of e.g. the Dirac algebra basis, the contractions have to be performed for each event and \texttt{CAMORRA} will produce helicity amplitudes at a lower rate. Finally there are the massive spinors, for which the construction policy is determined by the \texttt{spin\_vector\_type} of the model. In particular, this class implements a function constructing for a given massive momentum $p$ the spacelike spin vector $s(p)$ for which $p\cdot s=0$ and $s^2=-1$. The package contains three implementations, which carry the names introduced in \cite{AndreevPRD62}:
\begin{eqnarray*}
\texttt{KS\_type}: s^{\mu}(p)&=&\frac{p^{\mu}}{m}-\frac{mk_0^{\mu}}{p\cdot k_0}\,,\\
\texttt{helicity\_type}: s^{\mu}(p)&=&\frac{p^0p^{\mu}-m^2\delta_0^{\mu}}{m|\vec{p}|}\,,\\
\texttt{polarised\_type}: s^{\mu}(p)&=& \delta^{i\mu}+\frac{p^i}{m(p^0+m)}(p^{\mu}+m\delta_0^{\mu})\,.\\
\end{eqnarray*}
where $i$ denotes the beam direction. The massive spinor wave functions are then constructed according to the standard formula
\begin{equation*}
u_{\pm}(p)=\frac{(\slashed{p}+m)(1\pm\gamma^5\slashed{s}(p))}{2\sqrt{k_0\cdot(p+ms(p))}}u_{\mp}(k_0)\,.
\end{equation*}
and the antiparticle spinors by $v_{\lambda}(p)=C\bar{u}_{\lambda}^T(p)$. Again, the spinor formulas for the three choices above were explicitly coded, avoiding multiple contractions.
\subsection{Fermion flow reversal}
\begin{figure}
\includegraphics[width=\textwidth]{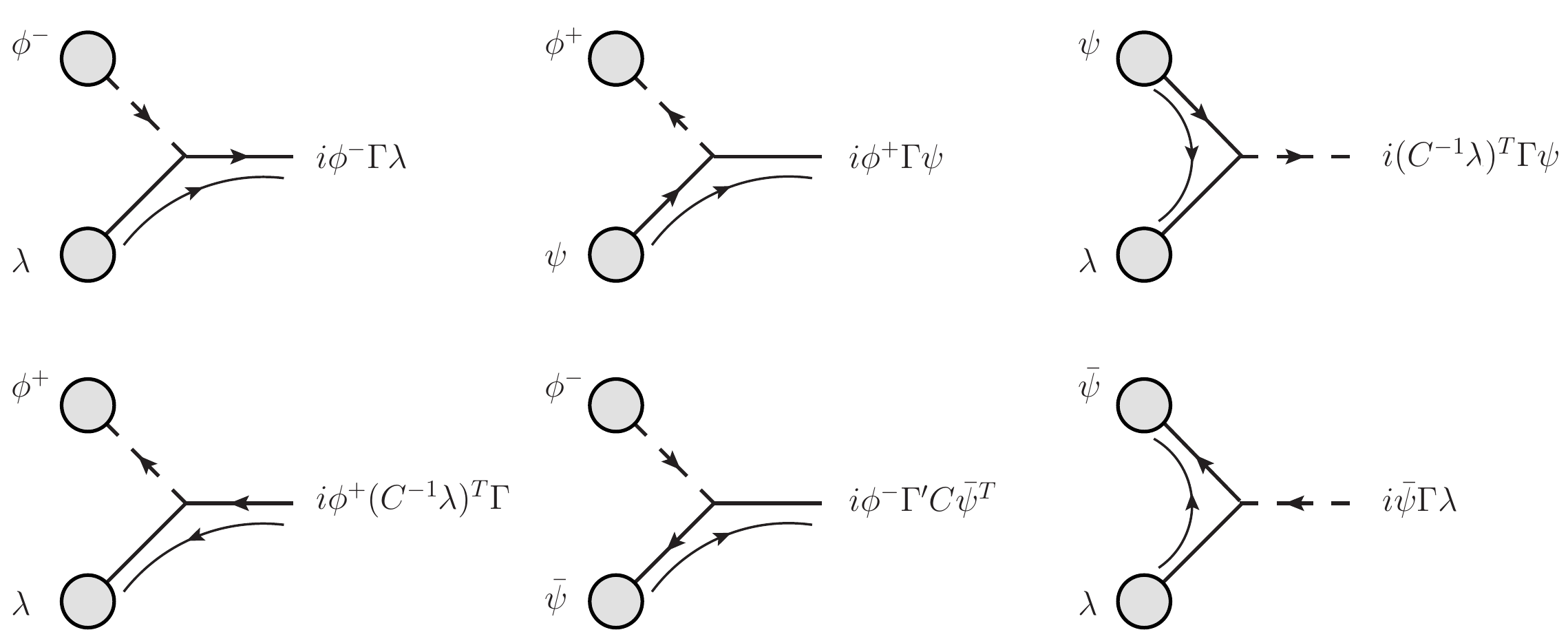}
\caption{Recursive relations for vertices containing one Dirac fermion (denoted $\psi$ or $\bar{\psi}$) and one Majorana fermion (denoted $\lambda$). The charged bosons are denoted by $phi^{\pm}$ and the vertex Feynman rule $\Gamma$ is an arbitrary combination of Dirac matrices. The matrix $\Gamma'$ is corresponding charge-conjugate matrix \cite{DennerNPB291}, which acts on the Clifford algebra basis by $(\gamma^{\mu})'=-\gamma^{\mu},(\gamma^5)'=\gamma^5,(\gamma^{\mu}\gamma^5)'=\gamma^{\mu}\gamma^5$ and $(\sigma^{\mu\nu})'=-\sigma^{\mu\nu}$.}
\label{Majvertices}
\end{figure}

In previous paragraph, we emphasized that the particle and anti-particle spinor phases should be consistent with the definition of the charge conjugation matrix. This is because \texttt{CAMORRA} computes scattering amplitudes containing Majorana currents by explicitly reversing fermion flows \cite{DennerNPB291} wherever it encounters a discontinuity. The strategy for treating Majorana fermions consistently in the recursive algorithm \cite{KleissARX2009} can be summarized as follows:
\begin{itemize}
\item External Majorana fermions are always treated as Dirac particles (never antiparticles), i.e. with the helicity spinors $u_{\lambda}(p)$ and $\bar{u}_{\lambda}(p)$ for resp. initial and final states.
\item At each vertex, the program assumes an input Majorana current to behave like a fermion, that is, a column spinor for outgoing and internal currents and a row spinor for incoming currents. For consistency \texttt{CAMORRA} therefore ensures that at any vertex a produced Majorana current behaves like a fermion as well.
\item Whenever a fermion-flow discontinuity is encountered, the program reverses the Majorana current by explicit charge conjugation. Therefore, models with no Majorana particles will never induce such reversals. When two combined Majorana currents give a fermion-flow violation, the first one will be reversed.
\end{itemize}
Just like the Fermi minus signs, the spinor reversal flags are determined at vertex tree construction. Note that a spinor reversal should not modify the current itself, but a copy of the subamplitude, since other vertex combinations--contributing to different Feynman graphs--might not require the current to be reversed. Therefore it was feasible to absorb the multiplication with the charge conjugation into the vertex contraction, resulting in recursive relations in e.g. fig. \ref{Majvertices}.
\subsection{Color flow optimization}
\begin{figure}
\begin{center}
\includegraphics[width=10 cm]{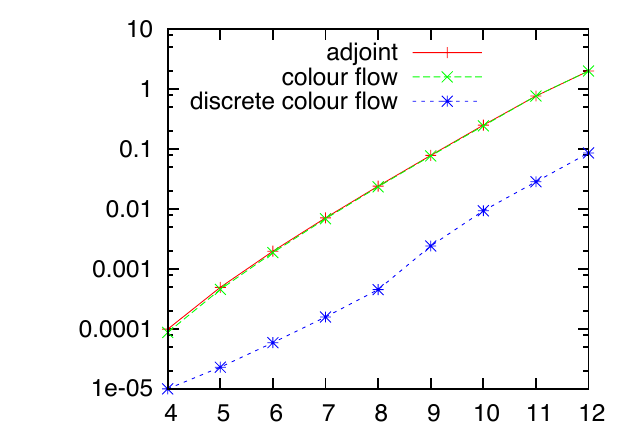}
\end{center}
\caption{Amplitude evaluation times (in s) for $n$ gluon amplitudes. Performance test was carried out by a 2.4 GHz intel core 2 duo processor.}
\label{speedtest}
\end{figure}

As already mentioned in section 2, \texttt{CAMORRA} accepts both adjoint and color-flow gluons and can run in either continuous or discrete color mode. In the \texttt{adjoint} color treatment, the recursive relations are optimized by omitting contractions over zero-valued color components, resulting in a performance comparable to the \texttt{color\_flow} mode. For discrete colors however, the continuous color flows can be traced through the vertex tree, and contractions over color tensors may be replaced with color flow combination rules \cite{PapadopoulosEPJC50}, often referred to as \emph{color-dressed recursive relations} \cite{DuhrJHEP0608}. This increases the processing speed by an expected factor around 9, compared to the continuous case (see fig. \ref{speedtest}). 
Note that the recursive relations contain all explicit terms decoupling the (unphysical) trace part of the color-flow gluon currents
\begin{equation}
G_{\alpha\beta}^{\mu}=T^a_{\alpha\beta}G_a^{\mu}
\label{cfgluon}
\end{equation}
such that the resulting amplitude is exact to all orders in $1/N_c$. The recursive relations from the available color structures in terms of the basis above were derived using FORM \cite{FORM}, and are included in the package.
\subsection{Built-in generators}
Included in the library are basic Monte Carlo generators for the color, helicity and momentum degrees of freedom. For the phase space generation, the standard RAMBO algorithm \cite{KleissCPC40} for uniformly distributed massless or massive momenta are available; for example, given a \texttt{CM\_algorithm} object \texttt{myalgo}, one calls\\
\\
\indent\texttt{myalgo.set\_momentum\_generator<uniform\_massless\_ps,std::random>(\&Ecm);}\\
\\
to generate massless momenta with a total invariant mass stored in \texttt{Ecm}. The second template parameter, \texttt{std::random} represents a random number generator class, in this case a wrapper class for the standard library's \texttt{rand} function which we have defined in the \texttt{std} namespace. The pointer to the phase space generator can be recovered by calling the \texttt{get\_momentum\_generator} member function. Cuts can then be applied as follows:\\
\\
\indent\texttt{using phase\_space\_variable::eta;}\\
\indent\texttt{myalgo.get\_momentum\_generator()->insert\_upper\_cut<eta>(3,\&etamax);}\\
\\
The code above restricts the third external particle momentum (the first outgoing one) to the $\eta<\eta_{\max}$ range. It will not restrict the momentum generation itself, but rather throw a zero amplitude (without actual tree evaluation) whenever \texttt{myalgo} encounters an event which does not pass the phase space cuts. Phase space variables are accessed by the \texttt{value} function:\\
\\
\indent\texttt{double myeta=myalgo.get\_momentum\_generator()->value<eta>(4);}\\
\\
The helicity degrees of freedom are inserted similarly via the \texttt{set\_helicity\_generator<}\emph{type, random number generator}\texttt{>()} member function. For helicities only the class \texttt{uniform\_helicities} is available for the \emph{type} parameter. In the discrete helicity mode, this generator throws a dice, while in the continuous case it generates uniformly an angle $\phi$ to construct the wave functions
\begin{eqnarray*}
e^{i\phi}u_{+}(p)+e^{-i\phi}u_{-}(p)\,,&&\textrm{for spin-1/2 particles,}\\ 
e^{i\phi}\varepsilon_{+}(p)+e^{-i\phi}\varepsilon_{-}(p)\,,&&\textrm{for massless spin-1 particles,}\\
e^{i\phi}\varepsilon_{+}(p)+\varepsilon_0(p)+e^{-i\phi}\varepsilon_{-}(p)\,,&&\textrm{for massive spin-1 particles,}
\end{eqnarray*}
Finally, the color sampling is directed by the member function \texttt{set\_colour\_generator<}\emph{type, random number generator}\texttt{>()}. Again, the \texttt{uniform\_colours} type throws dice in the discrete color case and generates uniformly normalized complex color phases in the continuous case. More physically relevant are the \texttt{adjoint\_QCD} and \texttt{colour\_flow\_QCD} classes. In continuous color models, the first generates real, normalized octets for external particles in the adjoint representation of $SU(N)$ and complex normalized triplets for $SU(N)$ fundamental-representation objects, whereas the second generates arbitrary $SU(N)$ matrices for the gluons using eq. \ref{cfgluon}. In the discrete color mode however, the \texttt{colour\_flow\_QCD} type will generate constrained color configurations, conserving the number of colors and anti-colors of each type \cite{PapadopoulosEPJC50}. The generation proceeds by generating a \emph{color connection}, i.e. an ordering of the anti-colors in the process with respect to the colors. After generating the colors uniformly, the anti-colors are determined by their position in the connection. The above color generation works for QCD-like theories, which have a constant static number \texttt{N\_c} defined in its model class header that represents the number of QCD colors.
\subsection{Tests}
The core routines and implemented models of \texttt{CAMORRA} were extensively tested on an event-by-event basis. Because of the flexibility and control over the computation allowed by the program, many cross-checking has been performed. These can be performed by the user as well, by typing \texttt{make check} after building the library. The testing facility includes
\begin{itemize}
\item Fermion contraction routine checks by comparing results in the Dirac and Weyl bases for gamma matrices.
\item Helicity state wave functions checks for normalization, mutual orthogonality and transversality. Comparisons between spin-summed amplitudes for different spin vector construction policies.
\item Tree-building algorithm checks by counting diagrams and symmetry properties of amplitudes.
\item Model implementation checks (especially for the \texttt{EWSM} and \texttt{SM} classes) by comparing amplitudes in different gauges.
\item Recursive relation consistency checks (especially of the Majorana Feynman rules) by rotating the choice of the final current in the vertex tree.
\item Color structure checks by comparing corresponding adjoint and color-flow amplitudes. Color combinations in the discrete color flow mode are checked against the continuous case.
\item Direct checks of helicity amplitudes by alternative computations, such as \cite{KleissNPB241} and the vanishing Parke-Taylor amplitudes in QCD \cite{ParkePRL56,BerendsNPB306,CachazoJHEP04}.
\item Cross check between supersymmetric and ordinary QCD amplitudes \cite{KunsztNPB271}
\begin{equation*}
A^{6g}(k_1+,k_2+,k_3+,k_4-,k_5-,k_6+)=-\frac{\langle k_4k_5\rangle}{\langle k_4k_6\rangle}A^{4g2\tilde{g}}(k_1+,k_2+,k_3+,k_4-,k_5+,k_6+)
\end{equation*}
\end{itemize}
\section{Conclusion and outlook}
We have written a scattering amplitude computation library that combines a full flexibility in terms of model implementation with the efficient recursive algorithm. This makes \texttt{CAMORRA} suitable for (tree-level) matrix element corrections to new physics processes with complex final states. Because the setup of the program is completely modular, new interaction and particle types may be included without modifying existing code. In terms of speed, \texttt{CAMORRA} is competitive with similar programs, whereas for multi-leg amplitudes, it exceeds model-independent programs based on the diagrammatic approach. Furthermore \texttt{CAMORRA} does not make use of external libraries (other than the standard C++ one) or symbolic manipulation programs and allows a user with basic knowledge of the programming language to implement new models and compute helicity amplitudes.\\
\\
In the future, the library may be equipped with a suitable phase space integrator to yield a full, model-independent matrix element event generator. Such a device should simulate the peaking structure of the integrand efficiently to achieve a good accuracy in reasonable computation times. Recently it has been proposed that a backwards traversal of the vertex tree allows a mapping of the amplitude peaks onto phase space \cite{COMIX,KALEU}, using a multichannel at each vertex node. This would be the obvious strategy towards a generic phase space integrator for \texttt{CAMORRA}.\\
\\
Another extension of the library--which may be immediately implemented--is the construction of an MSSM model class, possibly with an interface to SUSY Les Houches Accord parameters files \cite{SLHA,SLHA2}. In principle all the ingredients (color and Lorentz structures, Majorana fermions) are available and tested in \texttt{CAMORRA}, reducing the implementation to a translation of the Feynman rules \cite{RosiekPRD41} to the correct \texttt{add\_vertex} statements in the MSSM class.\\
\\
Finally, interfaces with the \texttt{FeynRules} \cite{FEYNRULES} and \texttt{LanHep} \cite{LANHEP} packages may be feasible to compute helicity amplitudes for given Lagrangian densities, and which would also facilitate cross-checking the outcomes with other packages. This may require some new definitions of Lorentz or color structures and possibly the inclusion of higher-spin particles.
\bibliography{mybibliography}
\bibliographystyle{hieeetr}
\newpage
\appendix
\section{The phase space interface}\label{PSinterface}
The momentum, helicity and color degrees of freedom of an external particle in a process that is loaded and built by the algorithm class are contained in objects of type\\
\\
\indent\texttt{CM\_algorithm<model\_type,N\_in,N\_out>::phase\_space\_type}\\
\\
whose addresses are accessed by the \texttt{get\_phase\_space()} member function of the algorithm class. References to the particle momentum components can for example be obtained as follows:\\
\\
\indent\texttt{myalgo.get\_phase\_space(0)->momentum(0)=450;}\\
\\
This sets the time-component of the first particle's 4-momentum to 450 GeV. Input of helicities depends on the \texttt{continuous\_helicities} flag in the model class. If it set false, the phase space type contains a helicity integer that may be set as follows:\\
\\
\indent\texttt{myalgo.get\_phase\_space(2)->helicity()=-1;}\\
\\
Valid choices for the helicity are 0 for scalars, $\pm 1$ for fermions and massless vectors and $\{-1,0,1\}$ for massive vector bosons. In the continuous helicity case, complex floating-point numbers can be assigned as coefficients belonging to the helicity states, e.g.\\
\\
\indent\texttt{myalgo.get\_phase\_space(2)->helicity\_phase(-1)=std::polar(3.14);}\\
\\
The color insertion proceeds in essentially the same way. If the \texttt{continuous\_colours} flag is set false, the color integers are inserted by e.g.\\
\\
\indent\texttt{myalgo.get\_phase\_space(0)->colour(1)=2;}\\
\\
If the first particle in the process is a gluon, this line of code can be used for instance to set its anticolor to 2\footnote{Remember that the colors run from 0 to $N_c-1$.} (assuming the algorithm runs in color flow mode). Note that particles with a single color index should always have it assigned by \texttt{colour(0)=...} (quarks do not have an anticolor in \texttt{CAMORRA}, there is no need to set it to zero). In the continuous color mode, the phase space instances contain color vectors or matrices (or tensors) of complex floating-point numbers. Their entries can be accessed by the function \texttt{colour\_coeff} which takes a variadic argument list of integers:\\
\\
\indent\texttt{myalgo.get\_phase\_space(0)->colour\_coeff(1,2)=complex<}\emph{value type}\texttt{>(1,0)};\\
\\
is the continuous equivalent of the previous color assignment. Note that if the length of the argument list does not match the color rank of the field, the above code will result in a run-time error.
\section{Examples}
The first example checks the vanishing Parke-Taylor amplitudes for a single 12-gluon event in QCD. The program instantiates an algorithm object called \texttt{algo}. After loading the process and constructing the vertex tree, it prints the number of Feynman diagrams\footnote{Note that the 4-gluon vertex in the QCD class is replaced by an auxiliary antisymmetric tensor field}. Then it uses the flat phase space generator at a center-of-mass energy of 1 TeV and the QCD color-flow sampler to generate momentum and color configurations. At this point all the gluon polarization vectors are initialized to $\varepsilon_-(p_i)$, and the amplitude is seen to vanish at double precision. It will remain zero after changing the wave function of gluon 2 to $\varepsilon_-(p_2)+\varepsilon_+(p_2)$. After including a second positive-helicity term, the amplitude becomes nontrivial.
\begin{verbatim}
#include <Camorra/QCD.h>
#include <Camorra/CM_algorithm.h>
#include <Camorra/uniform_massless_ps.h>
#include <Camorra/QCD_cols.h>
#include <Camorra/rcarry.h>

using namespace Camorra;

int main()
{
    CM_algorithm<QCD,2,10>algo("g,g > g,g,g,g,g,g,g,g,g,g");
    algo.load();
    algo.construct();
    std::cout<<"Nr of Feynman diagrams:"<<algo.count_diagrams()<<std::endl;
    double Ecm=1000;
    algo.set_momentum_generator<uniform_massless_ps,rcarry>(&Ecm);
    algo.set_colour_generator<colour_flow_QCD,rcarry>();
    algo.generate();
    
    std::cout<<algo.evaluate()<<std::endl;
   
    algo.get_phase_space(2)->helicity_phase(1)=1.0; 
    algo.get_phase_space(2)->helicity_phase(-1)=1.0;
    std::cout<<algo.evaluate()<<std::endl;
    
    algo.get_phase_space(3)->helicity_phase(1)=1.0;
    algo.get_phase_space(3)->helicity_phase(-1)=1.0;
    std::cout<<algo.evaluate()<<std::endl;
}
\end{verbatim}
After compilation the program will produce the output
\begin{quote}
\begin{verbatim}
Nr of Feynman diagrams: 50363463150
(-1.2957e-24,-4.11328e-24)
(-9.30578e-25,1.16322e-24)
(-2.01376e-15,-6.89148e-14)
\end{verbatim}
\end{quote}
Note that to ensure portability we used the \texttt{rcarry} random number generator \cite{RCARRY}. For completeness, the generated colors and momenta producing this output are included in the package.\\
\\
The second example involves the implementation of new model class. The toy model is called \texttt{SMseesaw} and adds a TeV-scale Majorana electron-neutrino \texttt{~nu\_e} to the standard model class. The model header \texttt{SMseesaw.h} file reads\\
\begin{verbatim}
#ifndef SMSEESAW_H_
#define SMSEESAW_H_

#include <Camorra/SM_base.h>
#include <Camorra/Minkowski.h>
#include <Camorra/Pauli_basis.h>
#include <Camorra/helicity_type.h>
#include <Camorra/col_flow.h>

using namespace Camorra;

class SMseesaw: public SM_base<SMseesaw,double>
{
    public:
        
        typedef double value_type;
        typedef Minkowski_type spacetime_type;
        typedef Pauli_basis Dirac_algebra_type;
        typedef helicity_type spin_vector_type;
        typedef colour_flow colour_treatment;
	    
        static const unsigned dimension=4;
        static const unsigned N_c=3;
        static const int beam_direction=3;
        static const bool coloured=true;
        static const bool continuous_helicities=false;
        static const bool continuous_colours=false;

        static value_type M_N;

        SMseesaw();
};
#endif
\end{verbatim}
The new model class is derived from the \texttt{SM\_base} class to load all the standard-model particles before adding new physics. In the header declaration we see the usual compile-time type definitions and constant static enumerable values. Note the \texttt{N\_c} integer, which is necessary to define when deriving from \texttt{QCD\_base}, \texttt{SM\_base} or \texttt{susy\_QCD\_base} as it determines the number of colors in the QCD sector. Note also the static \texttt{M\_N} floating-point number, which will store Majorana neutrino mass in the source file \texttt{SMseesaw.cpp},
\begin{verbatim}
#include "seesaw.h"

using namespace Camorra;

const unsigned SMseesaw::dimension;
const unsigned SMseesaw::N_c;
const int SMseesaw::beam_direction;
const bool SMseesaw::coloured;
const bool SMseesaw::continuous_helicities;
const bool SMseesaw::continuous_colours;

SMseesaw::value_type SMseesaw::M_N=1000;

SMseesaw::SMseesaw()
{
    add_fermion("~nu_e",&M_N);

    const std::complex<double>*coupling=&SM_base<SMseesaw,double>::Wnee;   

    add_vertex<vffR>("W+","~nu_e","e-",coupling);
    add_vertex<vffR>("W-","e+","~nu_e",coupling);
}
\end{verbatim}
The file begins with the initializations of the static data members of the model class. In the constructor, it adds a color-singlet Majorana electron-neutrino with mass \texttt{M\_N} and width zero. Then it constructs a vertex between the electron, $W$ bosons and Majorana neutrino, with couplings equal to the $W\bar{\nu}_{\ell}\ell$-couplings in the base class. After compiling this class into an object file, one can compute its predicted scattering amplitudes, for example for the $d,\bar{u}\rightarrow u,\bar{d},e^-,e^-$ process:
\begin{verbatim}
#include "SMseesaw.h"
#include <Camorra/CM_algorithm.h>
#include <Camorra/uniform_massless_ps.h>
#include <Camorra/uniform_hels.h>
#include <Camorra/QCD_cols.h>
#include <Camorra/rcarry.h>

using namespace Camorra;

int main()
{
    CM_algorithm<SMseesaw,2,4>algo("d,ubar > u,dbar,e-,e-");
    algo.load();
    algo.construct();
    std::cout<<"Nr of Feynman diagrams:"<<algo.count_diagrams()<<std::endl;
    double Ecm=500;
    algo.set_momentum_generator<uniform_massless_ps,rcarry>(&Ecm);
    algo.set_colour_generator<colour_flow_QCD,rcarry>();
    algo.generate();   
    algo.sum_spins();
    std::cout<<algo.evaluate_spin_sum()<<std::endl;
}
\end{verbatim}
which counts the number of diagrams and computes the spin-summed amplitude of random phase space point at a 500 GeV center-of-mass energy of the incoming quarks. It should yield the outcome
\begin{quote}
\begin{verbatim}
Nr of Feynman diagrams:4
1.69203e-12
\end{verbatim}
\end{quote}
The momentum and color configuration yielding this amplitude is included in the file \texttt{example2.txt}, to be inserted manually in case \texttt{rcarry} turns out to give different results on the user's system. The examples above and the phase space information files are included in the \texttt{examples} directory of the package.
\section{Installation and system requirements}
\texttt{CAMORRA} can be built on UNIX-based systems that are equipped with GNU \texttt{make} and \texttt{libtool}. It does not require any additional packages other than the standard library. The installation is straightforward: after downloading the archive to a directory of choice, unpack and decompress it by typing\\
\\
\indent\texttt{tar -xzvf camorra-1.0.tar.gz}\\
\\
in a terminal window. Then, move into \texttt{CAMORRA}'s main directory and type\\
\\
\indent\texttt{./configure --prefix=<}\emph{install directory}\texttt{>}\\
\indent\texttt{make}\\
\indent\texttt{make check} (optional, takes about 1h)\\
\indent\texttt{make install}\\
\indent\texttt{cd examples;make} (optional)\\
\\
where \emph{install directory} denotes the folder where the headers and library will be copied to. The \texttt{--prefix=...} may be omitted by users with root privileges, in which case the library will be installed in \texttt{/usr/local/lib} and the headers in \texttt{/usr/local/include}. More detailed information kan be found in the README and INSTALL files. In the doc directory of the distribution one finds a html reference list of all relevant classes and functions, generated with \texttt{doxygen}.
\end{document}